\documentclass[11pt,english]{article}
\usepackage{graphicx}
\usepackage{xcolor}
\usepackage{stmaryrd}
\usepackage{amstext}
\usepackage{braket}
\usepackage{caption}
\usepackage{amsfonts}
\usepackage{amsmath,amssymb,epsfig}
\usepackage{amscd}
\usepackage{mathrsfs}
\usepackage[applemac]{inputenc}
\usepackage[english]{babel}
\usepackage{enumitem} 
\usepackage{subfigure}
\usepackage{hyperref}
\usepackage{blindtext}
\usepackage{verbatim}
\usepackage{cancel}
\usepackage{cases}
\usepackage{cite}
\usepackage{tensor}

\hypersetup{
	pdftitle={},%
	pdfauthor={},%
	pdfsubject={},
	pdfkeywords={},%
	colorlinks=true,%
	linkcolor=blue,%
	citecolor=crimson,%
	linktocpage=true,%
	pageanchor=true
}

\newcommand{\bit}{\begin{itemize}}
	\newcommand{\eit}{\end{itemize}}
\newcommand{\bd}{\begin{description}}
	\newcommand{\ed}{\end{description}}

\newcommand{\bc}{\begin{center}}
	\newcommand{\ec}{\end{center}}

\newcommand{\be}{\begin{equation}}
	\newcommand{\ee}{\end{equation}}
\newcommand{\bea}{\begin{eqnarray}}
	\newcommand{\eea}{\end{eqnarray}}
\newcommand{\bs}{\begin{subequations}}
	\newcommand{\es}{\end{subequations}}

\usepackage[normalem]{ulem}

\definecolor{crimson}{rgb}{0.7, 0.08, 0.24}
\newcommand*{\affaddr}[1]{#1}
\newcommand*{\affmark}[1][*]{\textsuperscript{#1}}

\def\beqa{\begin{eqnarray}}
	\def\eeqa{\end{eqnarray}}
\def\bean{\begin{eqnarray*}}
	\def\eean{\end{eqnarray*}}

\numberwithin{equation}{section}

\usepackage[normalem]{ulem}

\usepackage{titlesec}
\titleformat{\section}
{\normalfont\sffamily\Large\bfseries}
{\thesection}{1em}{}
\titleformat{\subsection}
{\normalfont\sffamily\large\bfseries}
{\thesubsection}{1em}{}
\titleformat{\subsubsection}
{\normalfont\sffamily\large\bfseries}
{\thesubsubsection}{1em}{}

\title{\textbf{\textsf{Constraining Regular Black Hole Models with the Effective Field Theory treatment of General Relativity}}\vspace{0.25cm}}

\author{
	\textsf{Sami Viollet\thanks{The author passed away prior to the submission of this manuscript. The manuscript was prepared for submission posthumously, with only minor corrections and no changes to its scientific content.}
    }\\
	\affaddr{\affmark[1]\textsf{Aix-Marseille Universit\'e, Universit\'e de Toulon, CNRS, CPT,}}\\
	\affaddr{\textsf{13288 Marseille, France}}\vspace{-0.5cm}
}

\begin{document}
	
	\begin{sffamily}
		\maketitle
		\begin{abstract}
			\noindent
			\textsf{In this paper, we present a method for using perturbative predictions from treating general relativity as an effective field theory to constrain non-perturbative approaches to quantum gravity in black hole contexts. We start by reviewing a known result from this framework, which estimates the first quantum corrections to the interaction energy of a static particle near a black hole. We then propose a way to test non-perturbative quantum black hole models --- specifically those formulated as modified Schwarzschild metrics, also known as regular black holes --- against this perturbative result. To illustrate, we apply our method to the Hayward metric, showing that it does not align with the perturbative prediction, and we identify the most general metric that can reproduce the expected result.}
		
			\vspace{0.75cm}
		\end{abstract}

		\vspace{-0.25cm}
	\end{sffamily}

    \begin{center}
    \textit{This paper is submitted in memory of Sami, whose friendship and scientific curiosity will be deeply missed.}
    \end{center}
	
\section*{Introduction}

The quantization of general relativity has long been recognized as particularly challenging, partly due to the theory's non-renormalizability. 't Hooft and Veltman were the first to quantize perturbations to the Minkowski metric \cite{tHooft:1974toh}, concluding that while pure gravity is renormalizable at one loop, it becomes non-renormalizable at this same order if matter is included. 
Later, Goroff and Sagnotti showed that pure gravity also becomes non-renormalizable at the two-loop order \cite{goroff1985quantum}. 
In retrospect, the non-renormalizability of general relativity no longer seems a very surprising result. 
The reason being that spacetime is no longer regarded as a fundamental notion, but one that emerges from more elementary concepts that are as yet unknown. 
Today, there exist several non-perturbative approaches to quantum gravity (loop quantum gravity \cite{Rovelli_2004}, string theory \cite{Polchinski:1998rq,Polchinski:1998rr}, causal set theory \cite{Surya:2019ndm}, causal triangulation \cite{Ambjorn:2012jv}, asymptotic safety \cite{Spina:2025wxb} etc.) that attempt to formalize this idea. 
It seems therefore well established that general relativity is only an effective theory. 
But how does this relate to its non-renormalizability? 
Essentially, studying high-energy processes using degrees of freedom, in this case the metric, that lose meaning at such scales will naturally lead to inconsistencies. This is not the first time that a non-renormalizable theory has been understood as an effective one, emerging from a more fundamental theory. Fermi's non-renormalizable theory, for instance, was an effective theory which can be derived from the more fundamental theory of the weak interaction. Another example is Chiral Perturbation Theory, also non-renormalizable, which serves as an effective theory emerging from QCD.

Although effective theories cannot be completely satisfactory since they are inconsistent at high energies, they do remain predictive at low energy. Similarly, general relativity can be treated perturbatively as an effective field theory, enabling a certain number of predictions about the quantum behavior of gravity at low energy (compared with Planck energy). For a recent and pedagogical review of the subject, see \cite{Donoghue:2017pgk,burgess2004quantum}. One of the major predictions obtained by this approach was the estimation of quantum corrections to Newton's potential in the low energy regime. It took many years and the efforts of several research teams \cite{kirilin2002quantum, akhundov1997gravitational, donoghue1994leading} to achieve the correct result \cite{ donoghue1994general}. 
As this result comes from the effective treatment of general relativity, it is expected to be independent of the high energy theory describing the degrees of freedom from which emerges the familiar notion of spacetime. Despite the fact that such a theory is still unknown, we do know that it must agree at low energies with the effective treatment of general relativity. Just as Einstein knew that his theory of general relativity had to reproduce Newton's theory at low curvature.

In this article, we focus on black hole physics. 
We will review how this perturbative approach yields predictions about the quantum properties of these objects at low energy, or equivalently at low curvature. 
We will then see how these predictions can be used to constrain metric models of quantum black holes. 
Indeed, there is a large number of quantum black hole models formulated as modifications of the Schwarzschild metric \cite{hayward2006formation, ansoldi2008spherical, Ashtekar:2023cod, platania2023black} (and references therein). 
These are non-perturbative models in the sense that they are supposed to be valid in Planckian curvature regimes, and they are regular models in the sense that they are free of singularities. 
And as we motivated earlier, they must reproduce in the low energy limit the perturbative predictions made by treating general relativity as an effective field theory. 
Several quantum black hole metrics claim to reproduce the same corrections to Newton's potential as the perturbative approach \cite{de2015effective, DeLorenzo:2015taa, ling2022modified, scardigli2023metric}. 
However, these studies use different coordinate systems and compare quantities with distinct physical meanings. The challenge lies in the fact that, in QFT, the concept of potential is no longer fundamental. 
Multiple definitions of potential exist, each corresponding to different physical scenarios, such as interactions with fields or static sources \cite{donoghue1994general, muzinich1995long, Frob:2021mpb, frob2016quantum, frob2017quantum}. 
Similarly, in general relativity, various methods can define a potential of interaction. 
Thus, given a quantum black hole metric, one can derive modified Newtonian potentials in different ways, depending on the specific physical situation considered. This paper aims to identify a physical scenario that can be described by both the perturbative approach and a non-perturbative black hole metric. Assuming the perturbative approach is reliable, this comparison will help test the consistency of non-perturbative models. Moreover, it can help resolve ambiguities appearing in the construction of such non-perturbative models. To illustrate this point, we will consider models inspired by Loop Quantum Gravity which, as shown in \cite{munch2023generic, amadei2023landscape}, involve ambiguous choices that affect low-energy predictions. Regardless of their construction, we know that they must reproduce the perturbative predictions of the EFT treatment of general relativity. We will see how confronting these models with the perturbative results can help refine their construction and reduce ambiguities.

The paper is organized as follows. In the first section we recall the definition of two main notions of potentials in QFT. The notion of scattering potential associated with the scattering of plane waves, and the notion of static potential associated with the interaction of static sources. In the second section, we will review how these quantities have been computed at one loop in the context of gravity by treating general relativity as an EFT. We will be paying particular attention to the way the gauge is fixed in these computations since, if we want to compare these results, it will be important to do so in the same gauge. In the third section we will discuss the fact that in classical general relativity it is much easier to find a quantity corresponding to a static interaction potential than to a scattering potential. We will write this quantity explicitly and compute it for the case where there are two static point sources, one of which is much heavier than the other. This will be equivalent to computing the interaction potential of a static test particle in a Schwarzschild metric. We will then write down the most general modified Schwarzschild metric so that this interaction potential matches the one presented in the second section. This metric will be written in the harmonic gauge and then in the gauge using the area radius as the radial coordinate, the latter being the most common choice. In the last section we will illustrate how this result can be compared with popular quantum black hole models, namely the Hayward metric (as well as some generalizations of this metric) and the polymer black hole models (inspired by LQG). We will see that this comparison allows to test which non-perturbative quantum black hole models are consistent at low energy with the perturbative predictions, and how the latter can help to constrain ambiguities in the construction of such non perturbative models.

\noindent We use the signature $(1,-1,-1,-1)$. We take the convention $c=1$ and keep $G$ and $\hbar$ explicit.

\tableofcontents

\section{Potentials in quantum field theory}

It is well known that in quantum field theory, the notions of force and interaction potential between particles are not fundamental, but emerge from more fundamental concepts such as fields. Depending on the physical situation of interest, there are several ways of deriving a quantity that can be interpreted as an interaction potential. In this section, we recall how to obtain two of these notions of potential. The first is the scattering potential, this is the most commonly studied one and is obtained from the computation of a scattering amplitude. This quantity is well described in most quantum field theory textbooks (see \cite{Coleman:2018mew} for instance. In the case of QED, the author derives the formula for the Coulomb potential in this way). In a second step we will introduce the notion of static potential, which is the quantity we will be interested in next, and which has been extensively studied in the case of QCD in order to study quark confinement \cite{bander1981theories}. 

\subsection{Scattering potential} \label{scattering potential}

In this subsection, we recall how in QFT a scattering potential can be derived from the computation of the scattering amplitude of quantum field excitations. In order to do so, we will need to review how, in non-relativistic quantum mechanics, it is possible to obtain the differential cross section of a particle scattered on a potential. As these are standard results found in most reference textbooks on the subject \cite{cohen1986quantum, Coleman:2018mew, peskin2018introduction}, we will focus on the physical considerations and approximations used, rather than on the technical details. We will then see how, in QFT, it is possible by an inverse reasoning to obtain the expression of the scattering potential from the differential cross section. 

Let us start by writing the Schr\"odinger equation in the following way
\begin{equation}\label{sch eq}
-\frac{\hbar^2}{2m} \vec \nabla ^2 \psi(\vec r) + V(\vec r) \psi(\vec r) = E \psi(\vec r) ,
\end{equation}
where $V(\vec{r})$ is the potential term assumed to be known, and going to $0$ in the limit $r \rightarrow +\infty$ rapidly enough. In the most general case the solution to such an equation can be written as follows
 \begin{equation} \label{general sol}
 \psi(\vec{r}) =  \psi_0(\vec{r}) + \frac{2m}{\hbar^2}\int d^3\vec{r'}G(\vec{r}-\vec{r'})V(\vec{r'}) \psi(\vec{r'}),
 \end{equation}
where
\begin{equation}
(\vec{\nabla^2}+k^2) \psi_0(\vec{r}) = 0 \quad \text{and} \quad (\vec{\nabla^2}+k^2) G(\vec{r}) = \delta(\vec{r}) \quad \text{with} \quad k^2=\frac{2mE}{\hbar^2}.
\end{equation}
We are interested in the situation of a plane wave propagating towards the center of the scattering potential and then being scattered by it. It is therefore natural to take $\psi_0(\vec{r}) =e^{i\vec{k_i}.\vec{r}} $ representing the incident plane wave with wave vector $\vec{k_i} = k \vec{u_i}$, $\vec{u_i}$ being the direction of propagation of this plane wave. On the other hand, we retain only the retarded Green's function corresponding to an outgoing scattered wave. It is further assumed that the potential is only significant in a region $L$, and we are interested in the form of the wave function for $r\gg L$. The general solution \eqref{general sol} therefore takes the following form \cite{cohen1986quantum}
\begin{equation} \label{approx sol}
 \psi(\vec{r}) = e^{i \vec{k_i}.\vec{r}} + \frac{ e^{i k r} }{r}f(\theta , \phi) \quad \text{with} \quad f(\theta,\phi) =- \frac{m}{2 \pi \hbar^2} \int d^3\vec{r}e^{-ik\vec{u}.\vec{r}}V(\vec{r}) \psi(\vec{r}) \quad \text{and} \quad \vec{u} = \frac{\vec{r}}{r}.
\end{equation}
As expected, the second term in the right hand side describes a spherical outgoing wave. This solution is not analytic since the function $f$ depends on $\psi$. One could replace this $\psi$ by the entire right hand side and repeat this procedure infinitely. From there the usual approximation, called the Born approximation, consists of taking into account only the lowest order term in $V$, leading to the following approximate solution
\begin{equation} \label{f}
f(\theta, \phi) = - \frac{m}{2 \pi \hbar^2}\int d^3\vec{r}e^{-ik(\vec{u}-\vec{u_i}).\vec{r}}V(\vec{r}).
\end{equation}
One can then express the differential cross section in terms of \eqref{f}. To do so, remember that the differential cross section $\sigma(\theta,\phi)$ is defined as
\begin{equation}
dn = \sigma(\theta,\phi) F_i d\Omega,
\end{equation}
where $F_i$ is the incident flux of particles and $dn$ is the number of particles per second scattered in the solid angle $d\Omega$ centered around $\theta$ and $\phi$. It is then possible to compute $dn$ and $F_i$ from the solution \eqref{approx sol}. One can conclude that the differential cross section is given by \cite{cohen1986quantum}
\begin{equation} \label{dcs}
\sigma(\theta, \phi)=|f(\theta,\phi)|^2 = \left|\frac{m}{2 \pi \hbar^2}\int d^3\vec{r}e^{-ik(\vec{u}-\vec{u_i}).\vec{r}}V(\vec{r})\right|^2.
\end{equation}
Let us now turn to the study of scattering processes in QFT. In this context, there is no quantity such as $V(r)$ known a priory. It is however possible to compute transition amplitudes between an initial and a final state of fields. This quantity is generally referred to as $\mathcal{M}$ and is obtain perturbatively as a sum of many terms $\mathcal{M} = \sum_i \mathcal{M}_i$, each of which is associated with a Feynman diagram. It turns out that the differential cross section of the scattering process considered can be reconstructed from this quantity using the following formula \cite{peskin2018introduction}
\begin{equation} \label{dcs qft}
\sigma(\theta, \phi)=\frac{|\mathcal{M}|^2}{64\pi^2E_{CM}^2},
\end{equation}
where $E_{CM}$ is the total energy in the center of mass. Considering the non-relativistic limit (i.e. that the momentum of the incident waves is non-relativistic), it is expected that the predictions made in this way should correspond to the non-relativistic reasoning presented above. By replacing the expression for the differential cross section \eqref{dcs} in the result \eqref{dcs} and then performing an inverse Fourier transform, one can define a notion of scattering potential in QFT given by the following expression
\begin{equation} \label{V born}
V(\vec r) = \frac{\hbar^2}{4mE_{CM}}\int e^{i\vec k. \vec r}|\mathcal{M}|d^3k . 
\end{equation}
The transition amplitude $\mathcal{M}$ is usually written as a power series of the coupling constant between the fields (e.g. the fine structure constant $\alpha$ in QED). Therefore, the scattering potential can be computed at different order with respect to the coupling constant. In QED, the dominant term corresponds to the Coulomb potential, one can perform the computation at next order in $\alpha$ and would fine quantum corrections \cite{Coleman:2018mew}. Let us conclude this section by pointing out that the scattering potential introduced here is an approximate quantity (since it is based on the Born approximation), and that it is associated with scattering processes. There is, however, no simple quantity in classical gravity describing the process of plane-wave scattering. It will therefore be complicated to find the analog of \ref{V born} in the case of gravity with the result we would obtain in classical general relativity. This is why, in the next subsection, we will look at another notion of interaction potential with a clear physical meaning and a simple classical analog in general relativity.

\subsection{Static potential} \label{static potential}

The potential presented in the previous section is relevant for scattering processes, i.e. for dynamical situations. 
We will now introduce a notion of static potential describing the interaction of two static point sources. 
The advantage of this quantity is that, in the case of general relativity, there is a simple quantity describing the exact same physical situation which can then be compared with the quantum corrected one. 
This notion of static potential has been widely used in QCD to study the confinement of quarks \cite{bander1981theories}. 
We will briefly present how this quantity can be derived and subsequently illustrate how it can be used in the case of massive QED. 
Let us start by noting $J$ the source term that describes the interacting objects. 
We will then denote $\ket{n}_J$ the eigenstates of the Hamiltonian containing a source term $H(J)$, associated with the eigenvalues $E_n(J)$. 
The vacuum state when the source is switched off is denoted $\ket{0}$ and has zero energy. 
We assume that the source is switched on at $t=-\frac{T}{2}$ and off at $t=\frac{T}{2}$. The transition amplitude from vacuum to vacuum between these two instants is evaluated as follows
\begin{equation}
\bra 0 e^{-\frac{i}{\hbar}H(J) T}\ket{0} = \frac{Z(J)}{Z(J=0)}.
\end{equation}
By introducing the identity, one has
\begin{equation} \label{static energy 1}
\frac{Z(J)}{Z(0)} = \sum_n \bra{0} e^{-\frac{i}{\hbar}H(J) T}  \ket{n}_{JJ}\bra{n} 0 \rangle = \sum_n |\langle 0 \ket{n}_J|^2e^{-\frac{i}{\hbar}E_n(J)T} . 
\end{equation}
One can then perform the change $T \rightarrow T(1+i\epsilon)$, with $\epsilon \ll 1 $ a regulator, and take the limit $T\rightarrow +\infty$. This enables to isolate the first term of the sum, by removing the regulator $\epsilon$ we obtain  
\begin{equation}\label{static energy}
E_0(J) = \lim_{T \rightarrow + \infty}\frac{i\hbar}{T}\log{\frac{Z(J)}{Z(0)}}.
\end{equation}
The quantity thus obtained corresponds to the minimum of energy due to the presence of the source $J$. If we choose static sources, this energy will be their self-energy as well as their energy of interaction. This formula is of great use in QCD, for example, since it shows that the energy of interaction between quarks contains a linear term depending on the distance separating them, responsible for confinement \cite{bander1981theories, peskin2018introduction}. In order to illustrate the usefulness of \eqref{static energy 1}, we will quickly review how it can be applied in the case of massive QED, i.e. for a massive vector field (cf \cite{Zee:2003mt} for more details). The action of a massive vector field $A_{\mu}$ coupled to a source $J_{\mu}$ is
\begin{align} \label{action MV}
	S[A_{\mu}]&=\int dx^4 \left(-\frac{1}{4}F^{\mu \nu}F_{\mu \nu}+\frac{1}{2}m^2A_{\mu}A^{\mu}+A_{\mu}J^{\mu}\right)
\end{align}
with $F_{\mu \nu}=\partial_{\mu}A_{\nu}-\partial_{\nu}A_{\mu}$. The partition function obtained from this action is 
\begin{equation}
Z(J)=\int DA_{\mu} e^{\frac{i}{\hbar}\int dx^4 \left(A^\mu B_{\mu \nu} A^{\nu} +A^{\mu}J_{\nu} \right) }
\end{equation}
where $B_{\mu \nu}=(\partial^2+m^2)\eta_{\mu \nu}-\partial_{\mu}\partial_{\nu}$ is the heat kernel directly obtained from \eqref{action MV}. It is possible to rewrite this Gaussian integral as follows
\begin{equation}\label{part func MV}
Z(J)=Z(0)e^{-\frac{i}{2\hbar}\int d^4xd^4y J^{\mu}(x)D_{\mu \nu}(x-y)J^{\nu}(y)}
\end{equation}
We introduced the field propagator $D_{\mu \nu}=B^{-1}_{\mu \nu}$, which in this case admits the following expression
\begin{equation} \label{propagator MV}
D_{\mu \nu}(x)=\int\frac{d^4k}{(2\pi)^4}\frac{k_{\mu}k_{\nu}/m^2-\eta_{\mu \nu}}{k^2-m^2}e^{ik^\mu x_\mu}.
\end{equation}
We then choose a source representing two static particles, one positively charged at the origin and the other negatively charged at $\vec{R}$, more formally one has
\begin{equation} \label{source MV}
J^{\mu}(x)=\left[\underbrace{e v^{\mu} \delta(\vec{x})}_{J^{\mu}_p(x)}-
\underbrace{e v^{\mu}\delta(\vec{x}-\vec{R})}_{J^{\mu}_n(x)} \right] \Pi(x^0)
  \qquad \text{with} \qquad v^{\mu}=(1,0,0,0).
\end{equation}
The first term on the right hand side represents the positive particle, and the second one the negative one. This source is switched on at $x^0=-\frac{T}{2}$ and off at $x^0=\frac{T}{2}$, hence the presence of the function $\Pi$ that equals $1$ when $-\frac{T}{2}\le x^0 \le \frac{T}{2}$ and 0 otherwise. 
By inserting \eqref{propagator MV} and \eqref{source MV} in \eqref{part func MV}, and by noticing that the current conservation implies $k_{\mu}J^{\mu} = 0$, one can show that
\begin{equation}\label{intermediar 1}
\frac{Z(J)}{Z(0)}=\exp\left(\frac{i}{2\hbar}\int d^4xd^4y\frac{d^4k}{(2\pi)^4}\frac{\eta_{\mu \nu}(J^{\mu}_p(x)J^{\nu}_p(y)+J^{\mu}_n(x)J^{\nu}_n(y)-J^{\mu}_p(x)J^{\nu}_n(y)-J^{\mu}_n(x)J^{\nu}_p(y)}{k^2-m^2}e^{ik^\mu(x_\mu-y_\mu)}\right).
\end{equation}
Using \eqref{static energy} would lead to a relation between the argument of this exponential and the energy introduced by the source \eqref{source MV}, namely the self energy of the two particles as well as there energy of interaction. Since we are only interested by the energy of interaction between the two sources, we will only take into account the two last term in the exponential, the two others representing the self energy of the sources. One can show that the part of \eqref{intermediar 1} we are interested in can be written as follows
\begin{equation}
\exp\left(-\frac{i}{2\hbar}\int \frac{d^4k}{(2\pi)^4}dxdy\frac{J^{\mu}_p(x)J_{n\mu}(y)+J^{\mu}_n(x)J_{p\mu}(y)}{k^2-m^2}e^{ik(x-y)}\right) = \exp \left(\frac{iTe^2}{8\pi\hbar |\vec{R}|}e^{-m|\vec{R}|}\right).
\end{equation}
One can finally deduce by using \eqref{static energy} that the energy of interaction of the sources is
\begin{equation}
V_{\text{int}} = - \frac{e^2}{4\pi |\vec{R}|}e^{-m|\vec{R}|} . 
\end{equation}
This is the Yukawa potential, which is the expected result for a massive vector field such as that describing pions, for example. Note also that in the limit $m \rightarrow 0$, we recover the Coulomb potential, which is not surprising since the photon field is a massless vector field. 

In this section, we have introduced two different notions of potential, a scattering potential and a static potential which is actually more an energy than a potential, we shall therefore refer to it in the following as static energy of interaction. In the next section, we will show how general relativity, although non renormalizable, can be treated as an effective field theory, and how these potentials can be computed in this case.

\section{Perturbative quantum gravity} \label{Perturbative quantum gravity}

We will now review some of the main steps of the effective field theory treatment of general relativity (for a more complete review of the topic, see \cite{Donoghue:2017pgk, burgess2004quantum}). We will recall what the quantum corrections to the Newton's potential are that this approach enables to predict. First, in the case of scattering phenomena \cite{donoghue1994general} and then in the case of interaction between static objects \cite{muzinich1995long,modanese1995potential}. We will see that these results allow us to deduce certain quantum properties of black holes in the low energy regime. We will be paying particular attention to the way in which the gauge is fixed in these computations, since our aim is to compare these results with those obtained by describing a black hole with a (modified) Schwarzschild metric.

\subsection{Linearized gravity in the harmonic gauge} \label{Linearized gravity in the harmonic gauge}

The first step in the perturbative quantization of general relativity is to linearize the theory. Let us start by recalling the formulation of linearized gravity, in which a massive scalar field minimally coupled to gravity is added. The Lagrangian density of the total system is
\begin{equation}\label{L}
\mathcal{L}=\mathcal{L}_{EH} + \mathcal{L}_{\phi} = \frac{2}{\kappa^{2}}\sqrt{-g}R + \sqrt{-g}\frac{1}{2}\left(g^{\mu\nu}\partial_{\mu}\phi \partial_{\nu}\phi-m^{2}\phi^{2}\right), \\
\end{equation}
where we introduced $\kappa^{2}=32\pi G$. We then linearize the metric $g_{\mu \nu}$ by considering a small perturbation $h_{\mu \nu}$, the graviton field, on top of the Minkowski spacetime $\eta_{\mu \nu}$
\begin{equation} \label{perturbed metric}
	g_{\mu\nu}(x)=\eta_{\mu\nu}+\kappa h_{\mu\nu}(x).
\end{equation}
We will now look at how the perturbed metric is transformed by a change of coordinates. Let us consider an infinitesimal coordinate transformation
\begin{equation}
	\label{gauge transfo}
	x_\mu \to x_\mu + \kappa \xi_\mu (x) \ .
\end{equation}
Under such a transformation, the graviton field transforms as 
\begin{equation}
	\label{graviton transfo}
	h_{\mu\nu}\to h_{\mu\nu}+\partial_{\mu}\xi_{\nu}(x)+\partial_{\nu}\xi_{\mu}(x) \ . 
\end{equation}
\noindent One can check that the Lagrangian \eqref{L} is invariant under such a transformation. Linearized gravity is therefore obviously a gauge theory, and quantizing it will naturally lead us to the usual difficulties encountered in quantizing gauge theories (i.e. the non-invertibility of the propagator of the gauge field, in this case of the graviton field). To avoid this problem, it is important to fix the gauge. In what follows, we will work in the so called harmonic gauge (or De Donder gauge, or Lorentz gauge) which can be written in the following ways
\begin{equation}\label{harmonic condi}
\Box x^{\mu} = 0 \Leftrightarrow \Gamma^{\alpha}_{\mu \nu}g^{\mu \nu} = 0.
\end{equation}
In the context of linearized gravity, i.e. by considering \eqref{perturbed metric}, this gauge fixing condition can be written as
\begin{equation} \label{harmonic condi lin}
\partial^{\nu}h_{\mu \nu} - \frac{1}{2}\partial_{\mu} h = 0 .
\end{equation}
One can notice that this is not a complete gauge fixing since it is still possible to perform gauge transformations of the form \eqref{gauge transfo} with $\xi_\mu(x)$ satisfying
\begin{equation}\label{residual gauge}
\Box \xi^\mu = \partial^\nu \partial_\nu \xi^\mu = 0.
\end{equation}
It is however possible to make this gauge unique by imposing appropriate boundary conditions on the metric (the situation is similar to the one in QED, where the Lorentz gauge is not a complete gauge fixing either, and one needs to impose fall off conditions on the potential $A^{\mu}$ to completely fix the gauge \footnote{The infinitesimal gauge transformation of the gauge potential in QED is $A^\mu \rightarrow A^\mu + \partial^{\mu}c$. The equivalent of the harmonic gauge in QED is the Lorentz gauge, which is $\partial_\mu A^\mu = 0 $, leaving a residual gauge transformation satisfying
\begin{equation}\label{residual gauge QED}
\Box c =0 .
\end{equation}	
However, in addition to the Lorentz gauge, if one requires the spacetime fall conditions necessary for the applicability of the uniqueness theorem presented in section \ref{Linearized gravity in the harmonic gauge}, then the only solution to \eqref{residual gauge QED} is $\partial^{\mu}c = 0$.}). 
This is due to a theorem that ensures the uniqueness of the solution to the harmonic equation $\Box \psi = 0$ when $\psi$ satisfies a number of fall off conditions both in space and time. 
In the rest of this subsection, we will review this theorem, and then see how it can be applied to our situation. 
The proof of this theorem will not be reviewed, but can be found in \cite{fock2015theory}. The theorem states that if a solution $\psi$ of the harmonic equation satisfies the following conditions:
\begin{enumerate}
\item[\textbf{(1)}] $\psi$ is bounded everywhere.
\item[\textbf{(2)}] In the limit $r=\sqrt{x^2+y^2+z^2}\rightarrow +\infty$ the function $\psi$ behaves as $\psi = \mathcal{O}\left(\frac{1}{r}\right)$ and $\partial_{\mu}\psi=\mathcal{O}\left(\frac{1}{r}\right)$ for all time $t$.
\item[\textbf{(3)}] There is no incoming waves, i.e. in the limit $r\rightarrow +\infty$ and for all time $t_0 = t+r$ the function $\psi$ satisfies $\frac{\partial (r\psi)}{\partial r} + \frac{\partial (r\psi)}{\partial t}=0$.
\end{enumerate}
then $\psi$ is unique and vanishes everywhere. In order to see how one can use this theorem to establish the uniqueness of the harmonic gauge, we start by considering two sets of harmonic coordinates $x^{\mu}$ and $x'^{\mu}$, satisfying \eqref{harmonic condi} and related by a global diffeomorphism noted
\begin{equation} \label{global change coordinates}
	x'^{\mu} = a^{\mu} + \Lambda^{\mu}_{\nu}x^{\nu} + \psi^{\mu}(x),
\end{equation}
where $a^{\mu}$ and $\Lambda^{\mu}_{\nu}$  parametrize respectively the global translations and the Lorentz transformations. With this notation, the metric transforms as follow
\begin{align} \label{metric transfo}
\nonumber
g'^{\alpha \beta} &= \frac{\partial x'^{\alpha}}{\partial x^{\mu}}\frac{\partial x'^{\beta}}{\partial x^{\nu}}g^{\mu \nu} \\
&=\left[\Lambda^{\alpha}_{\mu}\Lambda^{\beta}_{\nu}+ \left(\Lambda^{\alpha}_{\mu}\frac{\partial \psi^{\beta}}{\partial x^{\nu}}+\Lambda^{\beta}_{\nu}\frac{\partial \psi^{\alpha}}{\partial x^{\mu}}\right) +\frac{\partial \psi^{\alpha}}{\partial x^{\mu}}\frac{\partial \psi^{\beta}}{\partial x^{\nu}}\right]g^{\mu{\nu}} 
\end{align}
Since, by assumption, these two sets of coordinates satisfy \eqref{harmonic condi}, the functions $\psi^{\mu}$ also satisfy $\Box \psi^{\mu} = 0$. We will now assume fall off conditions on $x^\mu$ and on the metric. In the limit $r \rightarrow +\infty$:
\begin{enumerate}
\item[\textbf{(i)}] $x^{\mu}$ converge to a Cartesian set of coordinates as $x^\mu = \text{cartesian} + \mathcal{O}\left(\frac{1}{r}\right)$.
\item[\textbf{(ii)}] The metric converges to the Minkowski metric as $g_{\mu \nu} = \eta_{\mu \nu} + \mathcal{O}\left(\frac{1}{r}\right)$.
\item[\textbf{(iii)}] There are only outgoing gravitational waves (which is a reasonable assumption since we will be interested in isolated systems) i.e. $h_{\mu \nu}=\text{Outgoing waves}$.
\end{enumerate}
One can argue that if there exists another set of coordinates $x'^{\mu}$, related to $x^{\mu}$ by \eqref{global change coordinates} and also satisfying the properties \textbf{(i)}, \textbf{(ii)} and \textbf{(iii)} then $\psi^\mu$ must satisfy the conditions of the previous theorem and are equal to zero everywhere. First of all, since the coordinate transformation \eqref{global change coordinates} is a diffeomorphism, $\psi^{\mu}$ is necessarily bounded everywhere, thus satisfying condition \textbf{(1)} of the uniqueness theorem. On the other hand, one can note that according to \eqref{global change coordinates} if $x^{\mu}$ is a Cartesian set of coordinates at infinity, then $x'^{\mu}$ is also one if $\psi^{\mu}=\mathcal{O}\left(\frac{1}{r}\right)$ at infinity and for any time. In addition, we have assumed that the metric in the $x^{\mu}$ coordinates converges to the Minkowski metric at infinity in the following way
\begin{equation}
	g^{\mu \nu} - \eta^{\mu \nu} = h^{\mu \nu} = \mathcal{O}\left(\frac{1}{r}\right) \quad \text{when} \quad r\rightarrow +\infty.
\end{equation}
It follows from \eqref{metric transfo} that the metric in the $x'^{\mu}$ coordinates also converges in a similar way to the Minkowski metric if $\frac{\partial \psi^{\alpha}}{\partial x^{\mu}} = \mathcal{O}\left(\frac{1}{r}\right)$ at infinity and for all time. Hence condition \textbf{(2)} is satisfied. Finally, as in the $x^{\mu}$ coordinates, we want that 
\begin{equation}
g'^{\mu \nu} - \eta^{\mu \nu} = \text{Outgoing waves} \quad \text{when} \quad r\rightarrow +\infty.
\end{equation}
By using \eqref{metric transfo}, one can see that this condition imposes at leading order in $\frac{1}{r}$
\begin{align}
&\left(\Lambda^{\alpha}_{\mu}\frac{\partial \psi^{\beta}}{\partial x^{\nu}}+\Lambda^{\beta}_{\nu}\frac{\partial \psi^{\alpha}}{\partial x^{\mu}}\right)\eta^{\mu \nu} = \text{Outgoing waves} \\
&\Leftrightarrow \frac{\partial \psi^{\beta}}{\partial \tilde{x}^{\alpha}}+\frac{\partial \psi^{\alpha}}{\partial \tilde{x}^{\beta}}=\text{Outgoing waves},
\end{align}
where we introduced $\tilde{x}^{\alpha} = \Lambda ^{\alpha}_{\nu}x^{\nu}$. By using the following identity
\begin{equation}
\frac{\partial}{\partial x^{\beta}}\left(\frac{\partial \psi^{\alpha}}{\partial x ^{\nu}}+\frac{\partial \psi^{\nu}}{\partial x ^{\alpha}}\right) + \frac{\partial}{\partial x^{\alpha}}\left(\frac{\partial \psi^{\nu}}{\partial x ^{\beta}}+\frac{\partial \psi^{\beta}}{\partial x ^{\nu}}\right) + \frac{\partial}{\partial x^{\nu}}\left(\frac{\partial \psi^{\beta}}{\partial x ^{\alpha}}+\frac{\partial \psi^{\alpha}}{\partial x ^{\beta}}\right) = 2\frac{\partial^2 \psi^{\nu}}{\partial x^{\alpha} x^{\beta}}.
\end{equation} 
One can conclude that 
\begin{equation}
\frac{\partial^2 \psi^{\nu}}{\partial \tilde{x}^{\alpha} \tilde{x}^{\beta}} = \text{Outgoing waves} \Leftrightarrow \psi^{\nu} = \text{Outgoing waves},
\end{equation}
implying that condition \textbf{(3)} is also satisfied by the $\psi^{\mu}$. Therefore they satisfy all the conditions of applicability of the theorem of uniqueness and have to vanish everywhere. One can conclude that there is only one set of coordinates satisfying the harmonic gauge condition and the previous fall off conditions, up to global Poincar\'e transformations.

\subsection{Graviton correction to Newton's potential} \label{Graviton correction to Newton's potential}

We will now sketch out the perturbative quantization of gravity. This is an approach initiated long ago by t'Hooft and Veltmann in \cite{tHooft:1974toh}, a more pedagogical presentation can be found in \cite{veltman1981course}. 
The aim is to present the minimal formalism needed to understand how graviton corrections to Newton's potential can be obtained, first from scattering processes and then from static sources.  
The first step is to add a gauge fixing and a ghost term to the Einstein Hilbert action, leading to the following pure gravity action
\begin{equation}\label{pure gravity}
\mathcal{S}_{\text{PG}} = \int d^4x \left[ \underbrace{\sqrt{-g} \frac{2}{\kappa^2}R}_{\mathcal{L}_{\text{EH}}} + \underbrace{\frac{1}{\xi} \sqrt{-g}\left(\partial^{\nu}h_{\mu\nu}-\frac{1}{2}\partial_{\mu}h\right)^2}_{\mathcal{L}_{\text{gf}}} + \underbrace{\sqrt{-g}\eta^{*\mu}\partial^{\lambda}\partial_{\lambda}\eta_{\mu}}_{\mathcal{L}_{\text{ghost}}} \right],
\end{equation}
where $\xi$ is usually taken to be $1$, and $\eta$ is the ghost field, which is a fermionic vector field. One can then couple gravity to a matter source. As the aim is to compute the potential introduced in the subsections \ref{scattering potential} and \ref{static potential}, different source terms will have to be considered. 

In order to determine the scattering potential introduced in subsection \ref{scattering potential}, we start by adding to the action \eqref{pure gravity} the following Lagrangian density corresponding to two massive scalar fields that will be scattered
\begin{equation}
\mathcal{L}_{\text{scalar field}} = \frac{1}{2}\sqrt{-g}\sum_{i=1,2}\left(g^{\mu \nu}\partial_{\mu}\phi_i \partial_{\nu} \phi_i -m_i^2\phi_i^2\right).
\end{equation}
The two fields interact via exchanges of gravitons and the scattering amplitude of an plane wave of $\phi_1$ with one of $\phi_2$ can be perturbatively computed. The computation can in principle be done at any order in $G$ in the low energy regime, without the non renormalizability of general relativity being a problem \cite{Donoghue:2017pgk}. In practice, the computation of this scattering amplitude was performed at one loop (i.e. at order $G^2$), and it was then possible to estimate the first graviton correction to Newton's potential via the formula \eqref{V born}. After several attempts and auto-corrections \cite{kirilin2002quantum, akhundov1997gravitational, donoghue1994leading}, the established result is the following one \cite{ donoghue1994general}
\begin{equation} \label{V scattered}
	V_{\text{scat}}(R) = -\frac{Gm_1m_2}{R}\left(1+3\frac{G(m_1+m_2)}{R} + \frac{41}{10 \pi}\frac{G\hbar}{R^2}\right) + \mathcal{O}(G^3),
\end{equation}
where the sub-index ``scat'' refers to the fact that this quantity is related to a scattering process. 
This is the best known result in the effective field theory treatment of general relativity, and it is notably this result that is compared with non perturbative black hole models \cite{scardigli2023metric, ling2022modified, de2015effective}. 
However, this quantity is related to the scattering of plane waves, and therefore of completely delocalized objects. 
It is hard to conceive that this prediction can be directly used to describe the quantum behavior of black holes, which are localized objects. 
It is sometimes argued that by taking the limit $m_1\gg m_2$, the result \eqref{V scattered} describes the interaction of a test object (of mass $m_2$) with a black hole (of mass $m_1$). 
But this limiting case still corresponds to the interaction of two delocalized objects.

On the other hand, it is possible to use the notion of static energy of interaction, introduced in subsection \ref{static potential}, to determine the first graviton correction to the energy of interaction of two static and localized sources. To do so, we consider that the sources involved are described by the following energy momentum tensor
\begin{equation}\label{static source}
	T^{\mu \nu}(x) = m\delta_0^{\mu}\delta_0^{\nu}\delta^{(3)}(x-x_1) + M\delta_0^{\mu}\delta_0^{\nu}\delta^{(3)}(x-x_2) .
\end{equation}
This tensor corresponds to two static and point-like sources, one with mass $m$ at $x_1 = (t_1,\vec 0)$ and the other with mass $M$ at $x_2=(t_2,\vec R)$. This energy momentum tensor corresponds to adding the following term to \eqref{pure gravity}
\begin{equation}
S_{\text{static}}=-m\int^{T}_0dt\sqrt{g_{tt}(t,\vec x_1)}-M\int^{T}_0dt\sqrt{ g_{tt}(t,\vec x_2)}.
\end{equation}
The formula \eqref{static energy} can then be used to perturbatively compute the energy of interaction the sources considered. Again, the computation can theoretically be performed at any order in $G$ in the low energy regime. The first quantum correction was computed in \cite{muzinich1995long} and is given by
\begin{equation} \label{V static b}
	V_{\text{static}}-\frac{GmM}{R}\left(1-\frac{G(m+M)}{2R}+\frac{17G\hbar}{20\pi R^2}\right).
\end{equation}
As can be seen, the expressions \eqref{V scattered} and \eqref{V static b} differ. This should not be surprising since, as we have already mentioned, these two quantities describe completely different physical situations and therefore have no reason to coincide. Note also that in the classical limit $\hbar \rightarrow 0$, the two results do not coincide either. Indeed, the expressions \eqref{V scattered} and \eqref{V static b} have different post-Newtonian (PN) corrections. This should not worry us either, since there are also several notions of potential in general relativity, corresponding to different physical situations and containing different post-Newtonian corrections. In the next section, we will focus on static interactions in the regime where one of the first objects is much heavier than the second (i.e. $M \gg m$). In this case, the potential \eqref{V static b} becomes
\begin{equation} \label{V static}
	V_{\text{static}}=-\frac{GmM}{R}\left(1-\frac{GM}{2R}+\frac{17G\hbar}{20\pi R^2}\right) + \mathcal{O}(G^3).
\end{equation}
This system, of two static objects in gravitational interaction with one of the two objects much heavier than the second, corresponds classically to a static test particle in a Schwarzschild metric. In the following section, we will present the formula giving the energy of interaction of such a static particle in a Schwarzschild metric. We will then generalize the computation for a modified metric with quantum terms and see under which conditions the result \eqref{V static} is reproduced.

\section{Comparison with modified Schwarzschild spacetime}

In this section, we will start by deriving the formula for the energy of interaction of a static test particle in a Schwarzschild metric. In particular, it will be important to write this formula in the same gauge as that used in the subsection \ref{Linearized gravity in the harmonic gauge}. It will then be possible to observe that the formula thus obtained agrees with \eqref{V static} in the classical limit. We will then use a generalized version of this formula to show that the Hayward metric, which is one of the most common modified Schwarzschild metrics, does not allow us to recover \eqref{V static}. We will close the section by writing the most general modified metric that reproduces the perturbative result.

\subsection{Static potential in general relativity}

Let use consider a test particle of mass $m$, for now not necessarily static, evolving in a Schwarzschild metric of mass $M$. The Lagrangian of such a particle is
\begin{equation}\label{L free part}
	L=-m\frac{d \tau}{d t},
\end{equation}
where $\tau$ is the proper time of the test particle and $t$ is the proper time of a static observer at infinity. We then consider that the point particle is moving in a Schwarzschild geometry and following a radial trajectory. The proper time is given by the following expression 
\begin{equation}\label{sch line}
	d\tau^2=\left(1-\frac{2M}{r}\right)dt^2-\frac{dr^2}{1-\frac{2M}{r}}.
\end{equation}
We introduce the velocity $v=\frac{dr}{dt}$, and by using \eqref{sch line} the Lagrangian \ref{L free part} becomes
\begin{equation}
	L_{\text{Sch}}=-m\left[1-\frac{2M}{r}-\frac{v^2}{1-\frac{2M}{r}}\right]^{1/2} .
\end{equation}
The sub index ``Sch'' indicates that this is the expression of the Lagrangian for a particle in a Schwarzschild spacetime. The next step is to perform a Legendre transformation to obtain the Hamiltonian of this test particle. The conjugate moment is obtained directly from \eqref{L free part}, and the expression for the Hamiltonian follows from this.
\begin{equation} \label{H classic}
	p=\frac{\partial L}{\partial v}=m\frac{\frac{v}{1-\frac{2M}{r}}}{\left[1-\frac{2M}{r}-\frac{v^2}{1-\frac{2M}{r}}\right]^{1/2}} \quad \text{and} \quad H_{\text{Sch}}=m\frac{1-\frac{2MG}{r}}{\left[1-\frac{2MG}{r}-\frac{v(p)^2}{1-\frac{2MG}{r}}\right]^{1/2}}.
\end{equation}
Since we are interested in a situation where the interacting objects are static, we impose $v=0$. And since we want to compare this result with the one loop computation presented in subsection \ref{V static}, we can expand this Hamiltonian to order $G^2$
\begin{equation} \label{H classic static}
	H_{\text{Sch}} = m + V^{\text{Sch}}_{\text{int}} \quad \text{with} \quad V^{\text{Sch}}_{\text{int}} = -\frac{G Mm}{r}\left(1+\frac{G M}{2 r^2}\right) +\mathcal{O}(G^3)
\end{equation}
We directly recognize the usual Newtonian potential term in the expression of the energy of interaction, together with a term of order $G^2$ which is a PM correction. There is obviously no quantum corrections in \eqref{H classic static} since the Schwarzschild spacetime is purely classical. We will consider in what follows other black hole metrics, modified by quantum terms. In order to test which of these metrics are able to reproduce the correct quantum correction, we need to generalize the previous computation to the case of a general static and spherically symmetric spacetime, i.e. for metrics of the form
\begin{equation} \label{general le}
ds^2=g_{tt}dt^2+g_{rr}dr^2+r^2d\Omega^2.
\end{equation}
In this case, the Lagrangian of a particle moving only radially is $L=-m\sqrt{g_{tt}-g_{rr}v^2}$. The previous canonical analysis can be reproduced leading to the following Hamiltonian
\begin{equation}
H=m\frac{g_{tt}}{\sqrt{g_{tt}-g_{rr}v^2}}.
\end{equation}
One can then require the test particle to be static, and by removing the energy of mass one can deduce the generalized expression for the energy of interaction.
\begin{equation} \label{V classic static gen}
V_{\text{int}}=m(\sqrt{g_{tt}}-1).
\end{equation}
Before using this result for the modified Schwarzschild metric, we will see how to switch from the area radius gauge used in \eqref{general le} (and in most of the modified black hole models one can find in the literature) to the harmonic gauge used in section \ref{Perturbative quantum gravity}.

\subsection{The harmonic gauge for spherically symmetric and static spacetimes} \label{The harmonic gauge for spherically symmetric and static spacetime}

The energy of interaction \eqref{V static} was obtained by imposing the harmonic gauge and some additional conditions listed in the subsection \ref{Linearized gravity in the harmonic gauge}. The quantity $R$ appearing in \eqref{V static} is therefore the harmonic radial coordinate defined by
\begin{equation}
	R=\sqrt{X_1^2+X_2^2+X_3^2},
\end{equation}
where $X_1$, $X_2$ and $X_3$ satisfy \eqref{harmonic condi}. This is not a common gauge choice when working with modified black hole metrics. The area radius $r$ is often preferred. In order to compare \eqref{V static} with \eqref{H classic static} and \eqref{V classic static gen}, we shall rewrite these results in the harmonic gauge supplemented by the boundary conditions of subsection \ref{Linearized gravity in the harmonic gauge}. The analysis that follows is similar to what has been done in \cite{Belinfante:1962zza}. It can be shown that in a static, spherically symmetric spacetime, the harmonic gauge \eqref{harmonic condi} imposes the following condition between $R$ and $r$ \cite{weinberg1972gravitation}
\begin{equation} \label{relation r and R}
\frac{d}{dr}\left[r^2\sqrt{\left|\frac{g_{tt}}{g_{rr}}\right|}\frac{dR}{dr}\right]-2R\sqrt{|g_{tt}g_{rr}|} = 0.
\end{equation}
Let us start by considering the case of the Schwarzschild metric. In this situation, the previous equation admits the following solution
\begin{equation}\label{sol R}
R(r)=C_1\left(r-MG\right)+C_2\left(\frac{1}{2}\left(r-m_1G\right)\log \left|\frac{1}{1-\frac{2MG}{r}}\right|-m_1G\right).
\end{equation}
This solution contains two constants of integration $C_1$ and $C_2$, as \eqref{relation r and R} is a second order differential equation. The presence of these two constants, which we are free to choose as we wish, reveals that the harmonic gauge is not a complete gauge fixing. We can impose the same fall off conditions on $R$ that we considered in subsection \ref{Linearized gravity in the harmonic gauge} to completely fix the gauge. On the one hand, condition \textbf{(1)} requiring harmonic coordinates to be bounded everywhere imposes that $C_2=0$. On the other hand according to condition \textbf{(2)}, the harmonic coordinates must coincide with the Cartesian coordinates at infinity. This is equivalent to requiring that $R(r) \xrightarrow{r \rightarrow + \infty } r$, in other words, imposing $C_1=1$. With these choices the solution \eqref{sol R} simplifies and becomes unique, the Schwarzschild line element can be written in the harmonic gauge as follows
\begin{equation} \label{harm Sch}
ds^2 = \frac{R-MG}{R+MG}dt^2 - \frac{R+MG}{R-MG}dR^2 + (R+MG)^2 d\Omega^2 \quad \text{with} \quad R=r-MG.
\end{equation}
In this gauge, the result \eqref{H classic static} becomes 
\begin{equation} \label{V sch harm}
V^{\text{Schw}}_\text{int}=-\frac{G Mm}{R}\left(1 - \frac{G M}{2 R}\right) +\mathcal{O}(G^3).
\end{equation}
As one can see, this result contains the same PN correction as \eqref{V static}. We can conclude that, in addition to having the same physical meaning, the two results \eqref{V sch harm} and \eqref{V static} coincide in the classical limit. In the next subsection, we will discuss the possibility of recovering the same quantum correction as in \eqref{V static} by choosing an appropriate modified Schwarzschild metric.

\subsection{The most general metric}

From the previous subsection, we established a method to test whether a regular black hole metric in the low-energy regime is consistent with the perturbative computation made by treating general relativity as an effective field theory. In this section, we take this analysis a step further by deriving the most general metric that reproduces this perturbative prediction. To achieve this, we first determine the metric such that inserting its $g_{tt}$ component, expressed in the harmonic gauge, into \eqref{V classic static gen} and expanding to order $G^2$, yields a result identical to \eqref{V static}. This is realized by the following expression
\begin{equation} \label{condi metric} 
	g_{tt} = 1-\frac{2 G M}{R}+\frac{ 2 G^2 M^2 }{ R^2}-\frac{ 17 G^2 \hbar M}{10 \pi R^3}+\mathcal{O}\left(G^3\right). 
\end{equation}
This represents the most general metric capable of reproducing the perturbative prediction \eqref{V static}. It is important to emphasize that this result relies on the harmonic gauge (along with certain boundary conditions specified in subsection \ref{Linearized gravity in the harmonic gauge}), meaning that the $R$ coordinate is the harmonic radius. 

To express the condition \eqref{condi metric} in the radial gauge, which is the most commonly used gauge, it is necessary to solve equation \eqref{relation r and R}. This would establish a relationship between $R$ and $r$, allowing us to rewrite \eqref{condi metric} in terms of $r$. However, since no constraints are imposed on $g_{rr}$, the equation \eqref{relation r and R} remains implicit and cannot be solved in full generality. Nonetheless, our focus is on making predictions at order $G^2$. As established in section \ref{Perturbative quantum gravity}, one-loop results are of order $\hbar$. It is therefore natural to adopt the following ans\"atze for the solution of equation \eqref{relation r and R}
\begin{equation}
	r=R + GM + a_1\frac{G \hbar}{R} + a_2 \frac{G^2M\hbar}{R^2} + \mathcal{O}(G^3) + \mathcal{O}(\hbar^2), \label{r(R)} 
\end{equation}
One can invert this relationship to obtain the expression of $R$ in terms of $r$ at order $\mathcal{O}(G^3)$ and $\mathcal{O}(\hbar^2)$
\begin{equation}
	R=r - GM - a_1\frac{G \hbar}{R} - (a_1+a_2) \frac{G^2M\hbar}{R^2} + \mathcal{O}(G^3) + \mathcal{O}(\hbar^2). \label{R(r)}
\end{equation}
On the other hand, we will consider the following ansatzes for the $g_{tt}$ and $g_{rr}$ components of the metric expressed in terms of the radial coordinate
\begin{align}
	&g_{tt} = 1-\frac{2GM}{r} + b_1 \frac{G\hbar}{r^2}  + b_2 \frac{G^2M\hbar}{r^3}+ \mathcal{O}(G^3) + \mathcal{O}(\hbar^2), \label{gtt(r)} \\
	&g_{rr} = -\left(1+\frac{2GM}{r}  + c_1\frac{G\hbar}{r^2} + c_2 \frac{G^2M\hbar}{r^3}\right)+ \mathcal{O}(G^3) + \mathcal{O}(\hbar^2). \label{grr(r)}
\end{align}
Our goal is now to determine the coefficients $a_1$, $a_2$, $b_1$, $b_2$, $c_1$, and $c_2$ so that both \eqref{relation r and R} and \eqref{condi metric} are satisfied. We begin by substituting \eqref{r(R)} into \eqref{gtt(r)}, yielding
\begin{equation}
	g_{tt} = 1-\frac{(2MR-b_1\hbar)G}{R^2}+\frac{2G^2M^2}{R^2}+\frac{(2a_1-2b_1+b_2)G^2M\hbar}{R^3} + \mathcal{O}(G^3) + \mathcal{O}(\hbar^2).
\end{equation}
To ensure consistency with \eqref{condi metric}, we impose the conditions
\begin{equation} \label{1st conditions}
	b_1=0 \quad \text{and} \quad a_1=-\frac{17}{20\pi}-\frac{b_2}{2}.
\end{equation}
Next, by inserting \eqref{R(r)}, \eqref{gtt(r)}, and \eqref{grr(r)} into \eqref{relation r and R}, we obtain
\begin{align}
	\frac{d}{dr}\left[r^2\sqrt{\left|\frac{g_{tt}}{g_{rr}}\right|}\frac{dR}{dr}\right]-2R\sqrt{|g_{tt}g_{rr}|} =&-\frac{(\frac{17}{10\pi}+(c_1+b_2)) G\hbar}{r} \nonumber\\
	&-\frac{G^2 M \hbar (\frac{17}{10\pi}+\frac{1}{2}(5b_2-2c_1+c_2))}{r^2}+ \mathcal{O}(G^3) + \mathcal{O}(\hbar^2).
\end{align}
To make the right hand side vanish at order $G^2$ and $\hbar$, we impose
\begin{equation} \label{3rd conditions}
	c_1=-\frac{17}{10 \pi}-b_2 \quad \text{and} \quad c_2 =- \frac{34}{5\pi}-7b_2.
\end{equation}
In summary, from \eqref{1st conditions} and \eqref{3rd conditions}, all parameters introduced in this section can be expressed in terms of $a_2$ and $b_2$. The most general relationship between $r$ and $R$ satisfying \eqref{relation r and R} and \eqref{condi metric} is
\begin{equation} \label{sol r and R}
	r = R + GM - \left(\frac{17}{20\pi}+\frac{b_2}{2}\right)\frac{G\hbar}{R} + a_2\frac{G^2\hbar M}{R^2} + \mathcal{O}(G^3) + \mathcal{O}(\hbar^2),
\end{equation} 
and the most general metric components in the radial gauge satisfying \eqref{condi metric} are
\begin{align} \label{metric area}
	&g_{tt}=1-\frac{2GM}{r} - b_2\frac{G^2 M \hbar}{r^3} + \mathcal{O}(G^3) + \mathcal{O}(\hbar^2) \\
	&g_{rr}=-\left(1+\frac{2GM}{r} + \frac{4G^2M^2}{r^2} - \left(\frac{17}{10 \pi}+b_2\right)\frac{G\hbar}{r^2} -\left(\frac{34}{5\pi}+7b_2\right)\frac{G^2\hbar M}{r^3}\right) + \mathcal{O}(G^3) + \mathcal{O}(\hbar^2)
\end{align}
The relation between $r$ and $R$ given by \eqref{sol r and R} satisfies the boundary conditions described in subsection \ref{The harmonic gauge for spherically symmetric and static spacetime}, making it the unique solution at this order in $G$ and $\hbar$. Additionally, the dependence of both \eqref{sol r and R} and \eqref{metric area} on two free parameters, $a_2$ and $b_2$, aligns with the fact that in the harmonic gauge, only $g_{tt}$ is constrained by \eqref{condi metric}, while $g_{RR}$ remains unconstrained. Indeed, at order $G^2$ and $\hbar$, $g_{RR}$ can be written as 
\begin{equation}
	g_{RR} = -\left(1+\frac{2GM}{R} + \frac{4G^2M^2}{R^2} + d_1 \frac{G \hbar}{R^2} + d_2d \frac{G^2 M \hbar}{R^3} \right) + \mathcal{O}(G^3) + \mathcal{O}(\hbar^2),
\end{equation}
which depends on two additional free parameters, $d_1$ and $d_2$. To reproduce the modified metric \eqref{condi metric} from the perturbative approach, both $g_{tt}$ and $g_{rr}$ in the radial gauge must be adjusted accordingly. Finally, we note that the relations \eqref{metric area} imply
\begin{equation} \label{condi 2}
	g_{tt} \ne -\frac{1}{g_{rr}}.
\end{equation}
In the next section, we will examine several modified black hole models proposed in the literature and evaluate their ability to reproduce the result \eqref{condi metric}. 

\section{Testing modified Schwarzschild metrics with the perturbative approach}

We now have everything we need to test which non perturbative quantum black hole models correctly reproduce the perturbative result presented in subsection \ref{Graviton correction to Newton's potential}. We will also see that this perturbative prediction can be helpful in constructing such non perturbative models. To do this, we simply have to write down the metric we want to test in the harmonic gauge and compare its $g_{tt}$ component with \eqref{condi metric}. Or, and it will often be more convenient, we can write the metric in the area radius gauge and compare its $g_{tt}$ and $g_{rr}$ components with \eqref{metric area}. Let us illustrate this procedure with two different models of modified Schwarzschild metrics.

\subsection{Hayward metric}

Let us start with the Hayward metric \cite{hayward2006formation}, expressed as follows
\begin{equation} \label{metric Hayward}
	g_{\mu \nu} dx^{\mu} dx^{\nu} = \left(1-\frac{2MGr^2}{r^3+2MG^2\hbar}\right)dt^2-\frac{dr^2}{1-\frac{2MGr^2}{r^3+2MG^2\hbar}}-r^2d\Omega^2.
\end{equation}   
This metric is written in the area radius gauge and one can easily expand its $g_{tt}$ and $g_{rr}$ components at order $G^2$ and $\hbar$ 
\begin{align}
&g_{tt} = 1-\frac{2 M G}{r} + \mathcal{O}(G^3) + \mathcal{O}(\hbar^2) \\
&g_{rr} = -\left(1+\frac{2 M G}{r} + \frac{4 G^2 M ^2}{r^2} \right) + \mathcal{O}(G^3) + \mathcal{O}(\hbar^2).
\end{align}
It is straightforward to see that it is impossible to choose the free parameters appearing in \eqref{metric area} to obtain the previous expansion of the Hayward metric components. The Hayward metric is therefore not consistent with the perturbative computation presented in section \ref{Perturbative quantum gravity}. We could have reached the same conclusion by noting that this metric does not satisfy \eqref{condi 2}. We also note that it is not possible to modify only the $g_{tt}$ component of the Hayward metric to reproduce \eqref{metric area}, despite what was suggested in \cite{de2015effective}. This subsection clearly illustrates how the perturbative result \eqref{metric area} can be used to test the relevance of some non perturbative modified black hole models.

\subsection{Polymer quantum black holes}

Let us move on to another category of models called polymer black hole models. This type of model emerged about twenty years ago and is inspired by loop quantum gravity (see \cite{Ashtekar:2023cod} and references therein for some examples). There are a large number of models of this type, since their construction is based on several arbitrary choices for the regularization of the Hamiltonian formulation of black holes. Each model corresponds to a particular choice. One of these choices concerns the functions used to regularize the phase space variables, known as polymer functions. The formal construction of this regularization and the physical role of such a choice have been studied in \cite{munch2023generic}. Here we show how the conditions \eqref{metric area} can be used to constrain this ambiguity. As explicitly shown in \cite{munch2023generic}, the metric associated with this class of models can be expressed as follows
\begin{equation} \label{polymer}
\begin{split}
ds^2 =& \frac{1}{r}
\left(\frac{\lambda_2^2}{16 f_2\left(\lambda_2P_2(1/r)\right)^2} - \frac{3 K_1 \lambda_2}{2 f_2\left(\lambda_2P_2(1/r)\right)}\right)d\tau^2 \\
&- \frac{\lambda_2^2/r^2}{16 f_1'(\lambda_1P_1(1/r))^2 f_2\left(\lambda_2P_2(1/r)\right)^2 g_{\tau \tau}(1/r)}dr^2 - r^2d\Omega^2,
\end{split}
\end{equation}
where $K_1$ is a constant related to the black hole mass in a way we will see soon, $\lambda_1$ and $\lambda_2$ are quantum parameters proportional to $l_p^2$ and $l_P$ respectively, $P_1$ and $P_2$ are phase space variables, and $f_1$ and $f_2$ are the so called polymer functions. The classical regime is obtained by taking $\lambda_2=\lambda_1=0$, $f_1$ and $f_2$ as the identity and using the classical solutions of the equations of motion for $P_1$ and $P_2$ (cf. \cite{munch2023generic}). From there it is possible to show that \eqref{polymer} reduces to the usual Schwarzschild metric. On the other hand, the usual choice for $f_1$ and $f_2$ to construct polymer black hole models is the sine function (cf. \cite{BodendorferEffectivequantumextended}). However, in \cite{munch2023generic} the authors argue that the only natural condition to impose on these functions is that they be periodic, and the usual choice of the sine function is not justified. The authors also showed that in order for \eqref{polymer} to reproduce the classical Schwarzschild metric in the low curvature regime (i.e., when $r \rightarrow +\infty$), one must have 
\begin{equation} \label{condi sch}
\lim_{r \rightarrow +\infty} f_{1,2}(\lambda_{1,2}P_{1,2}(1/r)) = 0, \quad \lim_{r \rightarrow +\infty} f'_{1,2}(\lambda_{1,2}P_{1,2}(1/r)) = 1 \quad \text{and} \quad \lim_{r \rightarrow +\infty} f''_2(\lambda_2P_2(1/r)) = 0.
\end{equation}
We now want to expand the metric components of \eqref{polymer} in the low curvature regime, i.e. when $r\rightarrow +\infty$ (expanding the metric components in $1/r$ is equivalent to expanding them in $G$ or $\hbar$, since both cases correspond to the weak field regime). As shown in \cite{munch2023generic}, after imposing \eqref{condi sch}, the $g_{\tau \tau}$ component of \eqref{polymer} has the following series expansion
\begin{equation}\label{gtt polymer}
g_{\tau \tau}=\frac{ 1}{16 }\left(1-\frac{24K_1P'_2(0)}{r} - \frac{ f_2^{(3)}(0)P'_2(0)^2}{2r^2}\right)+O\left(\frac{1}{r^3}\right).
\end{equation}
It can be seen that the Schwarzschild result is indeed reproduced at the lowest non trivial order, with $12K_1P_2'(0) =M$ the mass of the black hole. Moreover, it was also shown in \cite{munch2023generic} that the $g_{rr}$ component admits the following series expansion
\begin{equation}\label{grr polymer}
g_{rr} = -1+\frac{2M}{r}+\mathcal{O}\left(\frac{1}{r^3}\right).
\end{equation}
It is not hard to see that \eqref{grr polymer} is compatible with \eqref{metric area} by taking $c_1=c_2=0$, and no conditions need to be imposed on the polymer functions. On the other hand, for \eqref{gtt polymer} to be compatible with \eqref{metric area} one must impose
\begin{equation}
f_2^{(3)}(0)=0.
\end{equation}
It is interesting to note that the sine function, which is the usual choice for the polymerization procedure, does not satisfy this condition. Remember that we cannot impose $P_2'(0)=0$, since the black hole mass is proportional to this quantity. We will end this section by insisting on the fact that the choice of the sine function as the polymerization function, besides being arbitrary, does not lead to predictions in agreement with the perturbative approach. This subsection clearly illustrates how the perturbative result \eqref{metric area} can be used to help in the construction of non-perturbative modified black hole models by fixing potential ambiguities.

\section{Conclusion}

Despite the non renormalizability of general relativity, treating it as an effective field theory in a perturbative way enables finite and reliable predictions at low energy (i.e.~in regimes where the curvature is small compared to the Planck curvature). 
Predictions made in this way are independent of the complete underlying quantum gravity theory, as yet unknown, and must therefore be reproduced by the latter at low energy --- in a similar way that general relativity had to be in agreement with Newton's theory of gravitation in the non-relativistic limit of low curvature.
In this paper, we reviewed two of the main predictions made by this perturbative approach, which are quantum corrections to the scattering potential and to the energy of interaction of two static sources. 
The question addressed here is how to use these results to constrain non perturbative quantum gravity theories. 
In the absence of such a complete theory, we have focused on the case of black holes. 
A large number of quantum black hole models in the literature are formulated as modified Schwarzschild metrics. 
Although the main interest of such models is to propose a scenario in the Planckian curvature region and address the singularity problem, they have to agree with perturbative predictions in the low curvature regime. 
We have seen that, when one source is much more massive than the other, the perturbative quantum correction to the interaction energy of two static sources can be interpreted as the correction to the interaction energy of a static test particle and a black hole. 
This computation provides a clear prediction of the quantum modifications of the Schwarzschild black hole in the low energy regime. 
We have also seen how, given a modified Schwarzschild metric, the energy of interaction of a static particle in such a metric is related to the expression of the metric components. 
It was then possible to write down the most general metric reproducing the correct expression for the energy of interaction. 
In order to compare modified Schwarzschild metrics with the perturbative predictions, we had to be careful to treat the gauge freedom in the same way in both cases. 
In the perturbative approach, it is necessary to fix the gauge in order to obtain the graviton propagator. 
This is usually done by imposing the harmonic gauge and a number of boundary conditions. We have seen how to fix the same gauge in the non perturbative approach.
We have also seen how this most general metric can be written in the radius gauge area. Although the results are obtained in a specific gauge, they are compared with other black hole metrics expressed in the same gauge, ensuring that the comparison is well defined and can be carried out rigorously. Finally, we concluded by showing that, in the case of the Hayward metric, the perturbative prediction is not reproduced, which calls into question the validity of this model. 
On the other hand, it was possible to use the perturbative results to fix certain ambiguities inherent in the construction of non perturbative models, here the polymer black hole models. In particular, we have seen that the usual choice for these ambiguities, i.e. the sine function for polymerization functions, does not allow us to recover the perturbative prediction. 

\section*{Acknowledgments}
The manuscript benefited from a careful reading by Markus Fr\"ob, whose comments prior to submission are gratefully acknowledged.

\bibliographystyle{unsrt}
\bibliography{biblio}

\end{document}